\documentclass[10pt,twocolumn,letterpaper]{article}

\usepackage[pagenumbers]{cvpr}      

\usepackage{graphicx}
\usepackage{amsmath}
\usepackage{amssymb}
\usepackage{booktabs}
\usepackage[utf8]{inputenc} 
\usepackage[T1]{fontenc}    
\usepackage{url}            
\usepackage{booktabs}       
\usepackage{amsfonts}       
\usepackage{nicefrac}       
\usepackage{microtype}      
\usepackage{xcolor}         

\usepackage{array}
\usepackage{booktabs} 
\usepackage{multirow}  
\usepackage{enumerate}
\usepackage{graphicx}
\usepackage{colortbl}
\usepackage{amsmath}
\usepackage{amssymb}
\usepackage{bbm}
\usepackage{wrapfig}
\usepackage{caption}

\newcommand{\parahead}[1]{\noindent\textbf{#1}:\ }

\newenvironment{packed_itemize}
{\begin{itemize}
    \vspace{-\topsep}
    \setlength{\itemsep}{1pt}
    \setlength{\parskip}{0pt}
    \setlength{\parsep}{0pt}
}{\end{itemize}}

\newcommand{\nolistbottomspace}{\vspace{-\topsep}}

\newcommand{\chair}{\ensuremath{\mathcal{C}}}
\newcommand{\body}{\ensuremath{\mathcal{B}}}
\newcommand{\joint}{\ensuremath{\mathbf{j}}}
\newcommand{\joints}{\ensuremath{\mathbf{J}}}
\newcommand{\bone}{\ensuremath{\mathbf{b}}}
\newcommand{\bones}{\ensuremath{\mathbf{B}}}
\newcommand{\axang}{\ensuremath{\mathbf{r}}}

\usepackage[pagebackref,breaklinks,colorlinks]{hyperref}

\def\cvprPaperID{11087} 
\def\confName{CVPR}
\def\confYear{2022}

\begin{document}

\title{Learning Body-Aware 3D Shape Generative Models}

\author{Bryce Blinn\textsuperscript{1},
Alexander Ding\textsuperscript{1},
R. Kenny Jones\textsuperscript{1},
Manolis Savva\textsuperscript{2},
Srinath Sridhar\textsuperscript{1},
Daniel Ritchie\textsuperscript{1}\\\\
\small{\textsuperscript{\rm 1}Brown University, Providence, RI 02906}\\
\small{\textsuperscript{\rm 2}Simon Fraser University, Burnaby, BC V5A 1S6}\\\\
{\tt\small bryce\_blinn@brown.edu},
{\tt\small alexander\_ding@brown.edu},
{\tt\small russell\_jones@brown.edu},
{\tt\small msavva@sfu.ca},\\
{\tt\small srinath\_sridhar@brown.edu},
{\tt\small daniel\_ritchie@brown.edu}
}


\maketitle
\begin{abstract}
    The shape of many objects in the built environment is dictated by their relationships to the human body: how will a person interact with this object?
Existing data-driven generative models of 3D shapes produce plausible objects but do not reason about the relationship of those objects to the human body.
In this paper, we learn \emph{body-aware} generative models of 3D shapes.
Specifically, we train generative models of chairs, an ubiquitous shape category, which can be conditioned on a given body shape or sitting pose.
The body-shape-conditioned models produce chairs which will be comfortable for a person with the given body shape; the pose-conditioned models produce chairs which accommodate the given sitting pose.
To train these models, we define a ``sitting pose matching'' metric and a novel ``sitting comfort'' metric.
Calculating these metrics requires an expensive optimization to sit the body into the chair, which is too slow to be used as a loss function for training a generative model.
Thus, we train neural networks to efficiently approximate these metrics.
We use our approach to train three body-aware generative shape models: a structured part-based generator, a point cloud generator, and an implicit surface generator.
In all cases, our approach produces models which adapt their output chair shapes to input human body specifications.

\end{abstract}  

\section{Introduction}
\label{sec:introduction}

\begin{figure}[t!]
    \centering
    \includegraphics[width=\linewidth]{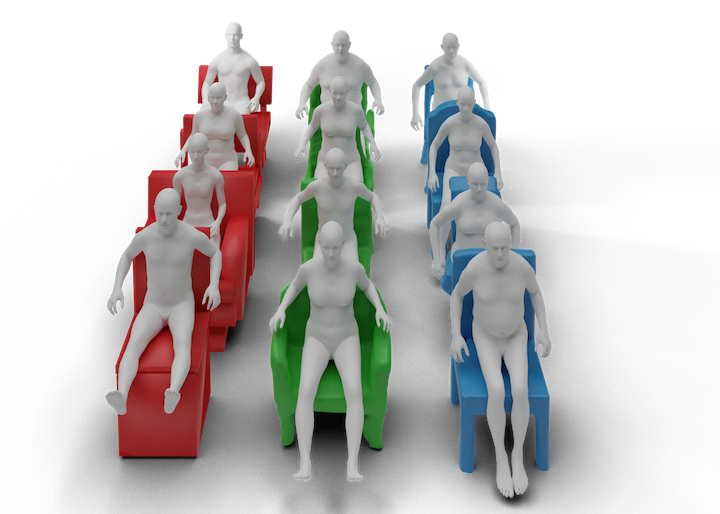}
    \caption{These chairs were generated by generative models fine-tuned by our method to produce chairs that will be comfortable for a given input body shape (shown sitting in each chair). Our method is general and can be applied to any latent-variable shape generative model. Here we show results from ShapeAssembly~\cite{13} (red), IM-Net~\cite{7} (green), and SP-GAN~\cite{6} (blue).
    }
    \label{fig:teaser}
\end{figure}

Why are 3D objects shaped the way they are? For many everyday objects, the answer is ``to allow people to interact with them'': tables support objects placed on top of them; doorknobs facilitate gripping and twisting; etc.
Technologies designed to assist in the creation of 3D objects should be aware of how people will interact with those objects.

Generative models of 3D shapes, particularly deep generative models, have received considerable recent attention in vision and graphics.
Such models can produce visually-plausible output geometries for a wide variety of object categories by mimicking patterns in a large training data.
This mimicry has limits, though: a visually-plausible object may fail to be functionally usable when a person tries to interact with it.
Consider the experience of many children making paper airplanes for the first time---many designs which look more or less like airplanes nevertheless do not fly!

We argue that the field of 3D shape generative modeling should move toward models that are \emph{explicitly} trained to produce outputs that are \emph{functional} when interacted with.
In this paper, we take a first step in this direction: we train generative models of 3D chairs that accommodate different body shapes and sitting postures.
Chairs are a good starting point for this investigation: they are ubiquitous in the real world, virtual 3D models of them are widely available, and the activity they afford (sitting) is relatively simple (compared to other activities that require dexterous manipulation).

We experiment with two types of body-aware chair generative models, each of which enforces a desirable property of its output chairs.
The first takes a body shape and produces chairs which will be comfortable for a person with that body shape; the second takes a sitting pose and produces chairs which accommodate that sitting pose.
To train these models, we define a ``sitting pose matching'' metric and a novel ``sitting comfort'' metric.
Calculating these metrics requires solving an optimization problem to sit the body into the chair, which is too computationally expensive to be used as a loss function for training our generative models.
Instead, we train neural networks to approximate the behavior of the pose matching and comfort metrics.

We apply our approach to training three types of chair generative models: a part-based generator, a point cloud generator, and an implicit surface generator.
In all cases, experiments show that our generative models successfully adapt their output chair shapes to the input body specification while still producing visually plausible results (see Figure~\ref{fig:teaser}).
In summary, our contributions are:
\begin{packed_itemize}
    \item A novel optimization procedure for chair sitting poses, as well as a physically-based metric of sitting comfort.
    \item Neural network proxy models which accurately approximate computationally-expensive functions that require optimization of bodies into sitting postures.
    \item A general approach for making any latent variable generative shape model body-aware.
\end{packed_itemize}
\section{Related Work}
\label{sec:related_work}

\parahead{Deep 3D Shape Generative Models}
Recent years have seen rapid advancement in deep neural networks for data-driven 3D shape generation.
Models have been proposed that synthesize shapes represented as volumetric occupancy grids~\cite{1,2,3}, point clouds~\cite{4,5,6}, implicit surfaces~\cite{7,8,9}, and others.
There are also structure-aware models which generating assemblies of primitives or other part geometries~\cite{10,11,12,13,14,15}.
We build upon three of these prior generative models: one generates objects as cuboid assemblies~\cite{13}, one as point clouds~\cite{6}, and one as implicit surfaces~\cite{7}.
We use the same methodology to make each type of generative model body-aware.

Our work is also related to recent work that fine-tunes an implicit shape generative model such that its outputs are physically connected and stable~\cite{Mezghanni_2021_CVPR}.
We extend this idea to human body awareness.
This prior system trains a neural network to serve as a stability loss function, instead of a more expensive physical simulation; we also train a neural networks to serve as loss function, instead of expensive optimization.
Our approach to fine-tuning generative models is also based on a learned warping of the latent space.
Our warping function must be body shape- and/or pose-conditional, though, which complicates the problem.

\parahead{3D Shape Functionality Analysis}
There is a considerable body of prior work on analyzing the \emph{functionality} of 3D objects, where ``functionality'' can be defined as the geometry of an object plus its interaction with other entities in a specific context~\cite{16}.
Prior work considering people as the `other entities' is related to our work.
SceneGrok learns a model which takes a 3D indoor scene as input and outputs a probability distribution over what human activities can be performed where~\cite{17}.
Follow-up work generated `interaction shapshots': localized arrangements of objects plus a person posed such as to be using those objects~\cite{18}.
In this work, scenes are generated by retrieving existing objects from a database; we seek to synthesize the geometry of objects themselves.
Pose2Shape takes an object's geometry as input and predicts a pose which a person might assume while interacting with that object~\cite{19}.
We focus on the inverse problem: taking a body shape or pose as input and producing an object which accommodates that body.
Finally, we develop a sitting comfort metric based on distributions of pressure exerted on the body.
A conceptually-related measure of sitting comfort has also been explored in prior work which analyzes human sitting behavior in videos~\cite{20}.

\parahead{Scene-Aware Body Generation}
There has been recent interest in the problem of generating human poses that fit a particular 3D scenes.
POSA~\cite{Hassan:CVPR:2021} uses a body-centric representation to place it at a static within 3D scenes, while SAMPL~\cite{hassan2021stochastic} places dynamic and plausible motions within 3D scenes.
Our goal is to solve the inverse problem: given human body poses, synthesize plausible 3D shapes.
\begin{figure*}[t!]
    \centering
    \includegraphics[width=\linewidth]{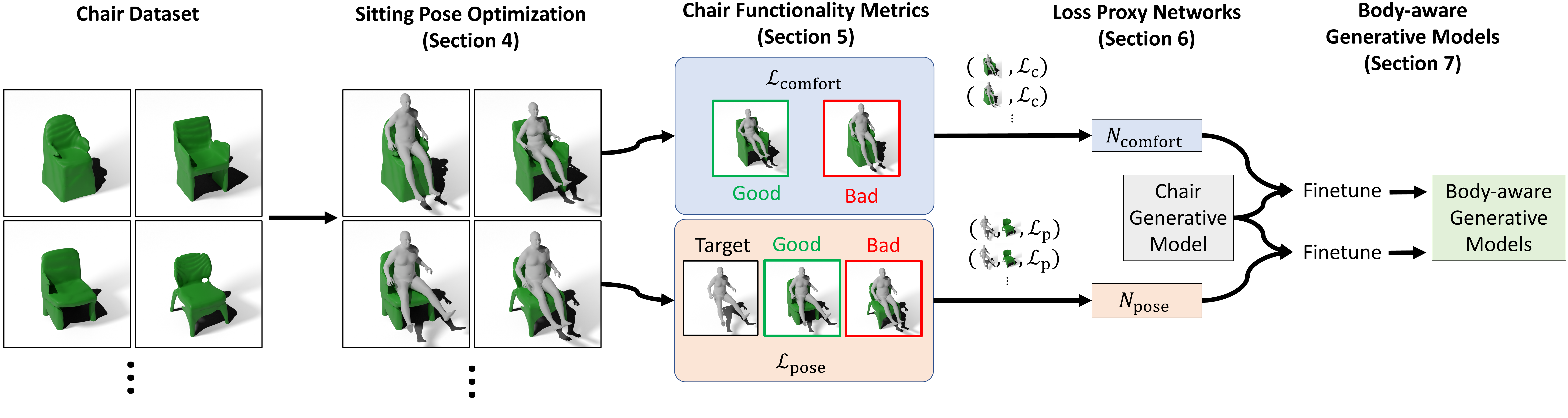}
    \vspace{-2em}
    \caption{Illustration of our pipeline. We first use physically-based optimization to compute sitting poses for a dataset of chairs. With these poses and chairs, we then compute two chair functionality metrics and construct sets of training samples that we use to learn neural loss proxies for the two metrics. These proxies allow efficiently finetuning of a chair generative model into a body-aware generative model.}
    \label{fig:overview}
\end{figure*}

\section{Overview}
\label{sec:overview}

Our objective is to train a 3D chair generative model that alters its output to accommodate an input human body shape or pose.
When given a body shape, it should produce output chairs in which a person with that body shape could comfortably sit.
When given a body pose, it should produce output chairs in which a person sitting would naturally assume that pose (or one similar to it).
We tackle the problem of creating such a generative model in four stages (see Figure~\ref{fig:overview}):

\parahead{Optimizing for sitting poses}
To train a generative model to produce chair shapes that accommodate specific bodies, we need a way to assess how well a given chair shape accommodates a given body. 
A prerequisite for this is the ability to simulate how a given body will sit in a given chair.
Thus, the first stage of our system is a physically-based optimization procedure which simulates a body settling into a chair.
Section~\ref{sec:pose} describes this procedure in more detail.

\parahead{Defining functionality metrics}
Given optimized sitting poses, we next define metrics which assess how well a chair meets our functional goals: being comfortable for a given body shape or supporting a given sitting pose.
Our sitting comfort metric is based on approximating the distribution of pressures exerted by the chair on the body.
Section~\ref{sec:obj} describes these metrics in more detail.

\parahead{Training loss proxy networks}
Given a body and a generated chair, evaluating the functionality metrics from the previous stage requires first optimizing for the sitting pose of the body in that chair.
This optimization makes our metrics prohibitively expensive for use as loss functions for training generative models.
Instead, we train efficient neural networks to approximate the behavior of these metrics.
See Section~\ref{sec:proxy} for more details.

\parahead{Training generative models}
Finally, we use the neural loss proxies to train body-aware 3D shape generative models.
We take a generative model pre-trained on a dataset of chair shapes, and adapt it to be body-aware by learning a body-conditional nonlinear warping of the pre-trained latent space.
Section~\ref{sec:generative} describes this approach in more detail.
\section{Sitting Pose Optimization}
\label{sec:pose}

Our goal is to define a procedure which takes a chair $\chair$ and a human body $\body$ and outputs a pose that the body would assume when sitting in the chair.
People can sit in the same chair in different ways and can shift continuously between different sitting postures.
In our work, we make the simplifying assumption that each body takes on a single, `relaxed' sitting pose in a given chair.
We define this pose as the solution to an energy minimization problem.
The energy function is comprised of terms that model physical constraints (e.g. gravity, collision), anatomical constraints (e.g joint angle limits), and characteristics of typical `relaxed' sitting poses (e.g. bilateral symmetry, large body/chair contact area).
This formulation is related to that of prior work in estimating sitting poses from images~\cite{21}.

\parahead{Human body model}
In our experiments, we use the SMPL human body model, a realistic 3D model of the human body based on skinning and blend shapes learned from thousands of 3D body scans~\cite{22}.
In SMPL, a body shape is specified by 16 parameters; a pose is specified by a global rigid translation, global rigid rotation, and one axis-angle rotation for each of 21 skeletal bones (a total of 69 parameters).
These pose parameters are our variables of optimization.

\parahead{Initialization}
We initialize the optimization by placing the body in a neutral sitting position (selected from the AMASS dataset~\cite{26}) suspended over the chair.
We set the pose's initial global rigid translation to $(0,\ y_{\text{max}} + 0.2,\ 0.5 (z_{\text{min}} + z_{\text{max}}))$, where $y_{\text{max}}$, $z_{\text{min}}$, and $z_{\text{max}}$ are parameters of the chair's bounding box (in a y-up coordinate system).

\parahead{Gravitational energy}
For the body to come to rest in the chair, it must minimize its gravitational potential energy.
To simplify the problem of modeling this energy, we represent the body as a connected assembly of rigid bodies, with mass concentrated at each joint $\joint$ in the body's skeleton.
We define the mass of each joint via a Monte Carlo point sampling of the body's interior volume: $m(\joint)$ is the percentage of sampled points which are closet to $\joint$, multiplied by a constant overall mass $M$ for the body (we use $M=1$ in our experiments).
The gravitational energy term is then
\begin{equation*}
    E_{\text{grav}}(\body, \chair) = \sum_{\mathbf{j} \in \joints(\body)} \mathbf{g} \cdot (j_y - y_{\text{min}}(\chair)) \cdot m(\mathbf{j}) \cdot M
\end{equation*}
where $\joints(\body)$ is the set of all body joints, $\mathbf{g}$ is the gravity vector, and $y_{\text{min}}(\chair)$ is the bottom of the chair's bounding box, i.e. the ground.

\parahead{Penetration energies}
We model contact of the body with the chair by minimizing a body/chair penetration energy:
\begin{equation*}
    E_{\text{pen}}(\body,\chair) = \sum_{\mathbf{v} \in \mathbf{V}(\body)} |\text{min}(d(\mathbf{v}, \chair),0)|
\end{equation*}
where $\mathbf{V}(\body)$ are the vertices of the body mesh and $d(\mathbf{x}, \chair)$ is the signed distance of a point $\mathbf{x}$ to the chair $\chair$.
We also include a term $E_{\text{self}}$ to penalize body self-penetrations, using an approach based on detecting colliding triangles using a bounding volume hierarchy and then evaluating local signed distance functions for each pair of colliding faces~\cite{21,24}.

\begin{figure*}[t!]
    \centering
    \setlength{\tabcolsep}{1pt}
    \begin{tabular}{ccccccc}
         Initial pose & No $E_{\text{sym}}$ & No $E_{\text{feas}}$ or $E_{\text{spine}}$ & No $E_{\text{zgrav}}$ & No $E_{\text{sit}}$ & All terms
         \\
         \includegraphics[width=0.16\linewidth]{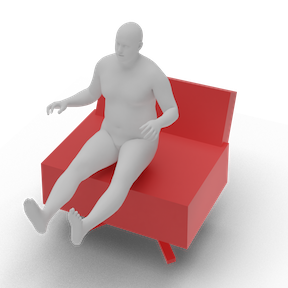}&
         \includegraphics[width=0.16\linewidth]{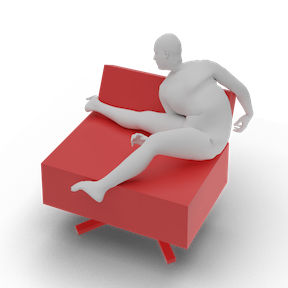} &
         \includegraphics[width=0.16\linewidth]{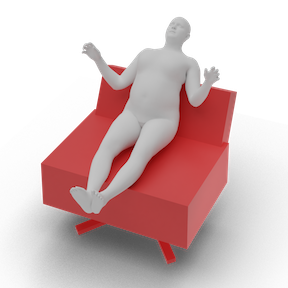} &
         \includegraphics[width=0.16\linewidth]{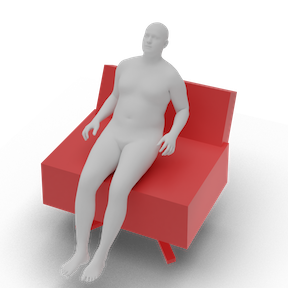} &
         \includegraphics[width=0.16\linewidth]{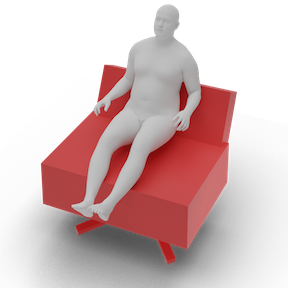}&
         \includegraphics[width=0.16\linewidth]{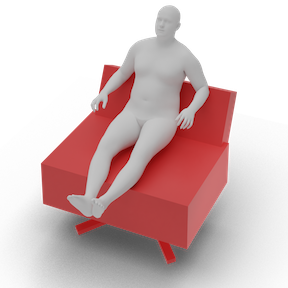}
    \end{tabular}
    \vspace{-1em}
    \caption{Demonstrating the effect of removing different energy terms from our sitting pose optimization objective.}
    \label{fig:sitting_opt_ablation}
\end{figure*}

\parahead{Pose feasibility energies}
Unconstrained optimization of SMPL joint angles can produce physically infeasible poses, as most joints in the human body have a limited range of motion.
Thus, we add an energy term which penalizes infeasible poses.
We adapt PosePrior, a binary classifier for determining whether a joint angle for a given bone in the SMPL skeleton is feasible~\cite{25}, by converting its outputs into a differentiable energy function.
For each bone in the body ($\bone \in \bones(\body)$), we uniformly sample its space of rotations $(\phi_\bone, \theta_\bone)$ and evaluate the classifier at each point, producing a binary validity grid $V_\bone$.
We then use a distance transform to produce a discrete field of signed distances to the boundary of the valid region $D_\bone$.
The pose feasibility energy is then
\begin{equation*}
    E_{\text{feas}}(\body) = \sum_{\bone \in \bones(\body)} \text{max} (\text{bilerp}(D_\bone, \phi_\bone, \theta_\bone) , 0)
\end{equation*}

As PosePrior does not include classifiers for spine joints, we use an additional energy term to penalize spine bending that deviates from that of the initial sitting pose:
\begin{equation*}
    E_{\text{spine}}(\body) = \sum_{\bone \in \bones^{\text{spine}}(\body)} || \axang_\bone - \axang^0_\bone ||_1
\end{equation*}
where $\axang_\bone$ is the axis-angle rotation vector for bone $\bone$.

Figure~\ref{fig:sitting_opt_ablation} illustrates the effect of removing these terms from the pose optimization.

\parahead{Sitting energies}
Optimizing the energy terms defined thus far will result in the body settling under gravity into a physically-feasible pose, but this pose may not resemble a sitting posture (e.g. the body may slide off of the chair and onto the ground).
Thus, we introduce terms which encourage the body to settle into a sitting pose.

The first term helps the body settle into a pose in which its back makes contact with the back of the chair (if one exists).
To achieve this goal, we introduce a gravitational energy term that acts along the z dimension (i.e. the front facing direction of the chair), pulling the body toward the backmost extent of the chair:
\begin{equation*}
    E_{\text{zgrav}}(\body, \chair) = \sum_{\joint \in \body} \mathbf{g}_z \cdot (j_z - z_{\text{min}}(\chair)) \cdot m(\mathbf{j}) \cdot M
\end{equation*}
where $\mathbf{g}_z$ is the gravity vector $\mathbf{g}$ rotated to be parallel to the z-axis.
Figure~\ref{fig:sitting_opt_ablation} shows the effect of omitting this term.

In addition, we include an energy term that encourages maximizing the contact area between the chair and the body's back and glutes.
If we let $\mathbf{V}^{\text{sit}}(\body)$ denote the set of body mesh vertices located on the back or glutes regions, then:
\begin{equation*}
    E_{\text{sit}}(\body,\chair) = \frac{1}{|\mathbf{V}^{\text{sit}}(\body)|} \sum_{\mathbf{v} \in \mathbf{V}^{\text{sit}}(\body)} \max(\text{d}(\mathbf{v}, \chair) - \tau, 0)
\end{equation*}
where $\tau$ is the distance threshold at which a vertex is considered in contact with the chair (we use $\tau = 0.001$).
Figure~\ref{fig:sitting_opt_ablation} illustrates the effect of removing this energy term.

\parahead{Bilateral symmetry energy}
As most human bodies are bilaterally symmetric, people typically assume bilaterally symmetric relaxed sitting poses.
Thus, our final energy term encourages the optimization to find such symmetric poses.
Let $\bones^{\text{sym}(\body)}$ be set the set of all symmetric pairs of joints in the body, $\mathbf{R}_{yz}$ be the reflection matrix about the $yz$ plane, $\mathbf{P}_{yz}$ be the projection matrix onto the $yz$ plane, and  $\mathbf{P}_{x}$ be the projection matrix onto the x axis. Then:
\begin{align*}
    E_{\text{sym}}(\body) = \sum_{(\bone_i,\bone_j) \in \bones^{\text{sym}}(\body)} ||\axang_{\bone_i} - \mathbf{R}_{yz} \axang_{\bone_j}||_1\\
    + \sum_{\bone \in \bones(\body) \setminus \bones^{\text{sym}}(\body)} || \mathbf{P}_{yz} \axang_b ||_1
    + ||\mathbf{P}_{yz} \axang_{\text{glob}}||_1 + ||\mathbf{P}_{x} \mathbf{t}_{\text{glob}}||_1
\end{align*}
The first term encourages all joints which have a symmetric twin to have the same rotation, up to reflectional symmetry.
The second term encourages all other joints not to rotate out of the $xy$ plane.
The third term encourages the body's global rotation to keep it within the $xy$ plane, and the fourth term encourages the body's global translation to do the same.

\parahead{Minimizing the total energy}
The total energy that we optimize is a weighted sum of all the energy terms defined above, which we minimize using the Adam optimizer~\cite{Adam}:
\begin{align*}
    &\alpha_{\text{grav}} E_{\text{grav}} + \alpha_{\text{pen}} E_{\text{pen}} + \alpha_{\text{self}} E_{\text{self}} + \alpha_{\text{feas}} E_{\text{feas}} + \\
    &\alpha_{\text{spine}} E_{\text{spine}} + \alpha_{\text{zgrav}} E_{\text{zgrav}} + \alpha_{\text{sit}} E_{\text{sit}} + \alpha_{\text{sym}} E_{\text{sym}} 
\end{align*}
We determine the weights of these terms empirically.
In our experiments: $\alpha_{\text{grav}} = 19.6,\ \alpha_{\text{self}} = 10,\ \alpha_{\text{feas}} = 0.1,\ \alpha_{\text{spine}} = 1,\ \alpha_{\text{zgrav}} = 9.8,\ \alpha_{\text{sym}} = 2.5$.
$\alpha_{\text{pen}}$ and $\alpha_{\text{sit}}$ vary by the shape representation used for chairs due to differences in the calculation of point-to-chair signed distances $d(\cdot, \chair)$.
For ShapeAssembly chairs, which use analytical point-to-cuboid signed distances, $\alpha_{\text{pen}} = 1$ and $\alpha_{\text{sit}} = 17$.
For IM-Net chairs, which use occupancy probability rather than a true signed distance, $\alpha_{\text{pen}} = 1.3$ and $\alpha_{\text{sit}} = 30$.
For SP-GAN chairs, which we convert to meshes using moving least squares~\cite{DeepMLS} and then to a discrete signed distance field via a distance transform, $\alpha_{\text{pen}} = 1.3 \cdot 10^{-2}$ and $\alpha_{\text{sit}} = 10^{-2}$.

\parahead{Running time}
Optimizing one sitting pose using this procedure takes upwards of a minute, on average, using our PyTorch implementation on an Intel i7-7800X machine with 32GB RAM and an NVIDIA GeForce RTX 2080Ti GPU.
GPU parallelism can accelerate the process (e.g. about 30 minutes for a batch of 200 chairs).
\section{Chair Functionality Metrics}
\label{sec:obj}

Now that we have defined a procedure for sitting a body on a chair, we can ask how well these sitting postures meet the functional goals we are interested in training generative models to satisfy: being comfortable for a given body shape or supporting a given sitting pose.
In this section, we define the metrics we use to quantify how well a given body $\body$'s sitting pose satisfies these goals for a given chair $\chair$.

\subsection{Comfort Loss}
\label{sec:comfort_metric}

We quantify the comfort of a body sitting in a chair by measuring the pressure the chair exerts on the body.
To compute these pressures, we need the equilibrium contact forces between the chair and the body; this requires solving for all forces acting upon the body when it is in static equilibrium.

\parahead{Solving for equilibrium forces}
Given a SMPL body mesh (simplified to 200 faces for computational efficiency), we discretize its interior volume into a tetrahedral mesh using fTetWild~\cite{fTetWild}.
Assuming our optimization has settled the body into the chair, we can assume that very little deformation to the body is required to achieve static equilibrium.
Thus, for simplicity, we treat the interior volume mesh as a connected assembly of rigid, constant-density tetrahedra.
We then compute equilibrium forces at each tet vertex by solving a system of linear equations with the following constraints:
\begin{packed_itemize}
    \item For each tetrahedron, the sum of the forces on its adjacent vertices plus the gravitational force acting upon it should be zero.
    \item For each tetrahedron, the net torque on each of its vertices about its center of mass should be zero.
    \item The sum of forces for all vertices in contact with the chair plus the gravitational force acting on the entire body should be zero.
    \item The net torque on each vertex in contact with the chair about the body's center of mass should be zero.
\nolistbottomspace
\end{packed_itemize}
The first two sets of constraints enforce a static equilibrium; the second two sets of constraints ensure that the equilibrium found is consistent with physical contact between the body and the chair.
In general, this system is over-determined (i.e. some small non-rigid deformation is necessary for equilibrium), so we solve it in least-squares sense.

\begin{figure}[t!]
    \centering
    \setlength{\tabcolsep}{1pt}
    \begin{tabular}{cccc}
        \includegraphics[width=0.24\linewidth]{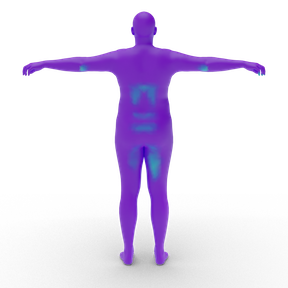} &
        \includegraphics[width=0.24\linewidth]{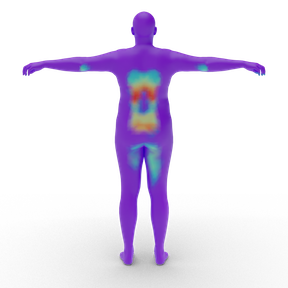} &
        \includegraphics[width=0.24\linewidth]{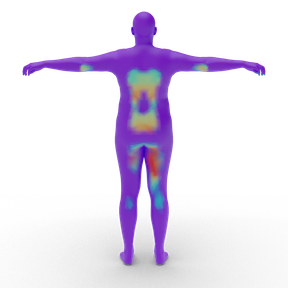} &
        \includegraphics[width=0.24\linewidth]{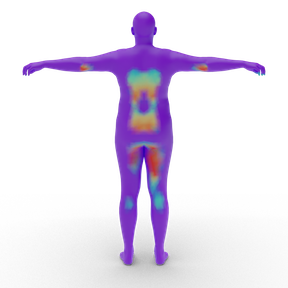}
    \end{tabular}
    \vspace{-1em}
    \caption{Visualizing the relationship between average pressure and comfort loss $\mathcal{L}_\text{comfort}$.
    We compute $\mathcal{L}_\text{comfort}$ for a large set of chairs paired with random body shapes and split the chairs into four bins based on these losses (left-to-right).
    In each image, we color each vertex by its average pressure across all bodies in that bin.
    }
    \label{fig:contact_viz}
\end{figure}

\parahead{Computing pressure}
Given these forces, we compute the total pressure  on the body as the sum of pressures on each triangular face $f$ of the body mesh in contact with the chair (i.e. the sum of normal forces divided by area):
\begin{equation*}
    \mathcal{L}_\text{comfort}(\chair,\body) = \sum_{f \in \body \cap \chair} \frac{\mathbf{F_v}(f) \cdot \mathbf{\hat{n}}(f)}{A(f)}
\end{equation*}
where $\mathbf{F_v}(f)$ is the sum of face $f$'s vertex forces, $\mathbf{\hat{n}}(f)$ is $f$'s surface normal, and $A(f)$ is $f$'s area.
Figure~\ref{fig:contact_viz} visualizes how the average pressures increase as $\mathcal{L}_\text{comfort}$ increases.

\subsection{Pose Matching Loss}

To measure how similar a sitting pose $S$ is to a desired target pose $T$, we simply use the sum of Euclidean distances between corresponding joints in the two poses:
\begin{equation*}
    \mathcal{L}_\text{pose}(\body_S, \body_T) = \sum_{i=1}^{|\joints(\body_S)|} || \joints(\body_S)_i - \joints(\body_T)_i ||_2
\end{equation*}
To make this metric invariant to global rigid translation, we remove the body root joint from the computation.
We also remove the hands and feet joints, as our sitting optimization has no energy terms governing their position.
\section{Loss Proxy Networks}
\label{sec:proxy}

Evaluating the functionality metrics defined above requires running expensive optimization to find a body's sitting pose in a chair.
It is not feasible to run this optimization on every iteration of training a chair generative model.
Instead, we train neural networks which approximate the behavior of the comfort and target pose loss functions.

\begin{figure*}[t!]
    \centering
    \includegraphics[width=\linewidth]{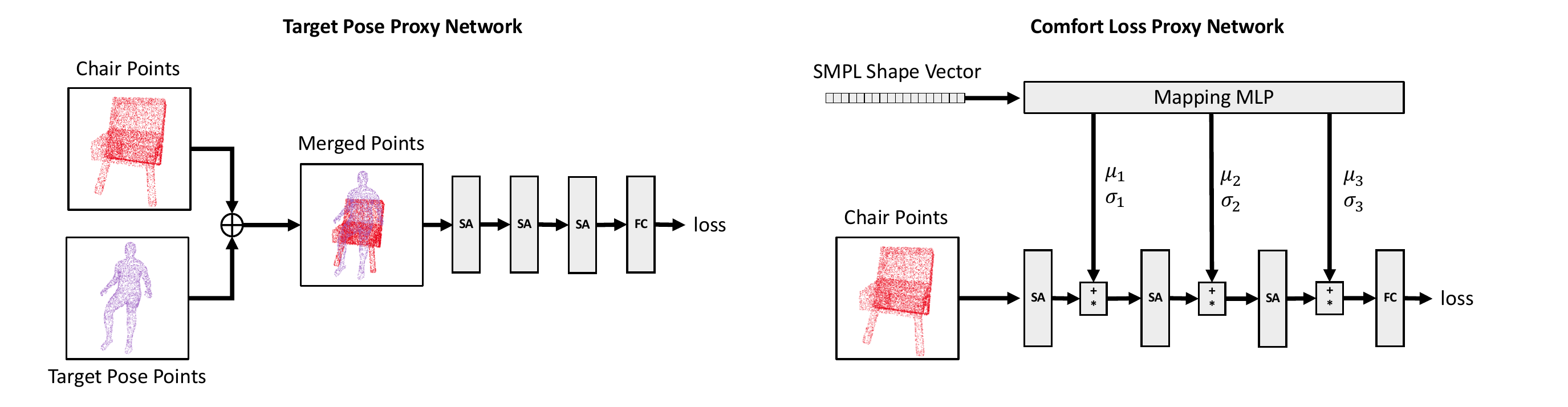}
    \vspace{-2em}
    \caption{Loss proxy network architectures. We use neural networks to approximate the two functionality losses we have introduced. Left: the target pose proxy network takes a merged point cloud of a chair and a target pose and passes it through a PointNet++ backbone (three set abstraction layers and a linear layer). Right: the comfort loss proxy network takes an 18-dimensional SMPL body shape vector which conditions the same PointNet++ backbone through Featurewise Linear Modulation (FiLM).}
    \label{fig:networks}
\end{figure*}

\subsection{Network Architectures}

Figure~\ref{fig:networks} shows the architectures of these proxy networks.
Both are based on PointNet++~\cite{28}: a point-sampled chair is passed through three PointNet++ set abstraction layers, after which the features are flattened and fed to a linear layer to produce the output loss value.
The networks differ in how they incorporate body conditioning information.
The pose loss network takes an additional point cloud depicting a body in the desired pose; the network should predict how close this pose will be to the pose the body would take when seated in the chair.
This point cloud is merged with the chair point cloud; each point is given an extra one-hot dimension indicating whether it is a chair or body point.
The comfort loss network takes as input only a body shape, specified as a 18-dimensional SMPL body shape feature vector $\mathbf{f}_\text{shape}$, the concatenation of the 16 body parameters as well as a one-hot encoding of the sex.
The network is conditioned on this vector via Featurewise Linear Modulation (FiLM)~\cite{29}, i.e. a MLP takes $\mathbf{f}_\text{shape}$ as input and outputs point-wise scale and shift parameters for each set abstraction layer.

\subsection{Training}

Given a dataset of chair shapes, we use the optimization procedure from Section~\ref{sec:pose} to optimize sitting poses for 100 different body shapes per chair, where body shape vectors $\mathbf{f}_\text{shape}$ are randomly sampled from a multivariate normal distribution fit to the the AMASS dataset~\cite{26}.
To produce training data for the comfort loss proxy network, we evaluate the comfort loss on all of these optimized (chair, body shape, pose) tuples, resulting in $100N$ (chair, body shape, comfort loss) training examples for a chair dataset of size $N$.
To produce $100N$ training examples for the target pose loss proxy network, for each chair, we evaluate the target pose loss on 100 different poses.
One of these is the optimized pose for the chair itself (which incurs zero target pose loss); the other 99 are randomly sampled from the set of optimized poses for other chairs.
Finally, this data is split 95\%/5\% into train/test.
Both networks are trained to minimize the absolute difference between their predicted loss and the ground truth.
To stabilize training, the ground-truth loss values are whitened (normalized to zero mean and unit variance).
See the supplement for details.
\section{Body-aware Generative Models}
\label{sec:generative}

\begin{figure}[t!]
    \centering
    \includegraphics[width=\linewidth]{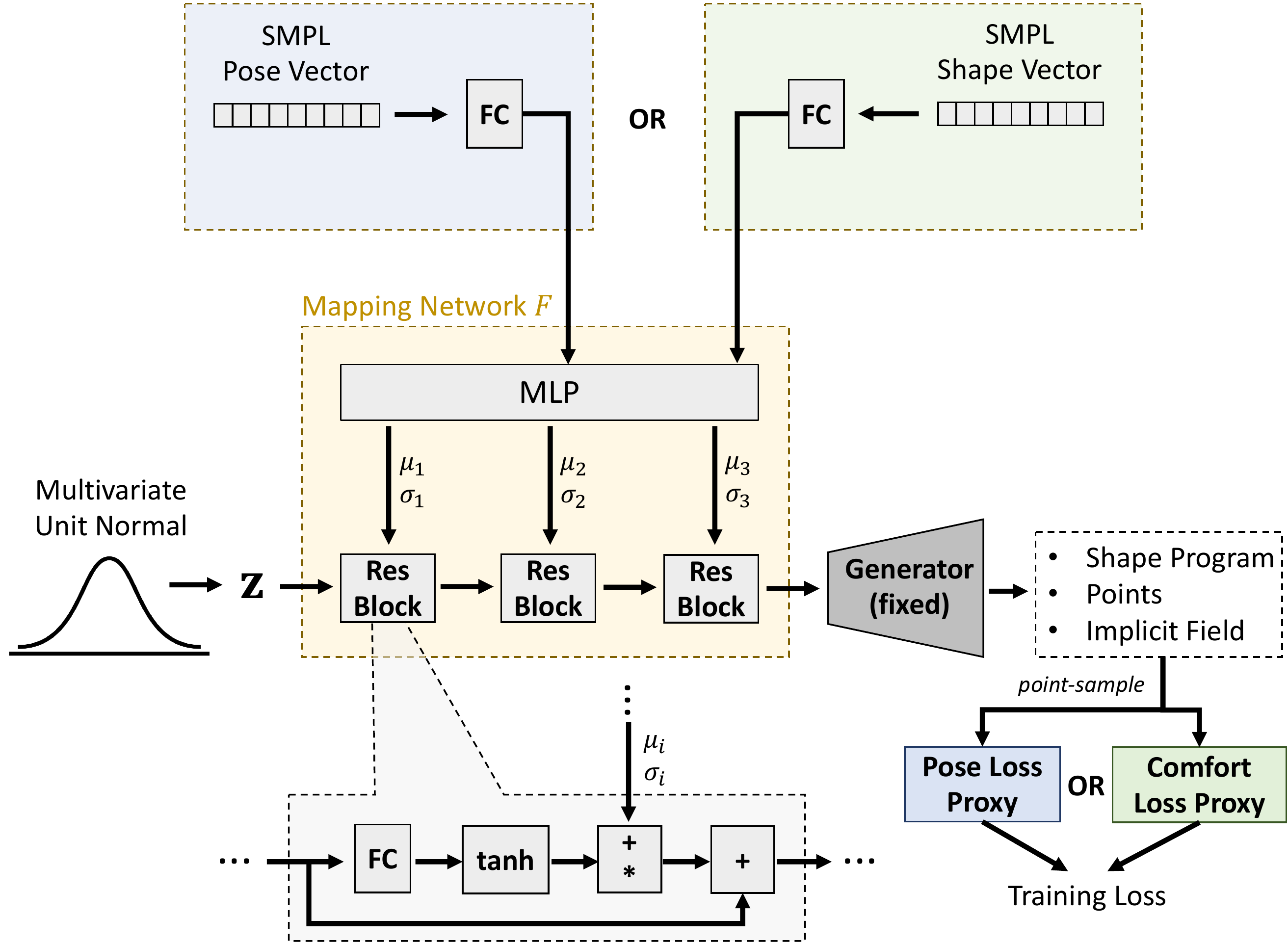}
    \vspace{-1.5em}
    \caption{Generative model architecture diagram.
    We use a mapping network $F$ to warp the latent space of a pre-trained shape generative model, conditioned on either a body shape or pose.
    Bell curve icon by Davo Sime from the Noun Project.
    }
    \label{fig:gen_condition}
\end{figure}

Finally, we use the proxy networks to train body-conditional generative models by fine-tuning a generative model pre-trained on a chair dataset.
Rather than fine-tune the generator (which could result in shapes that satisfy the proxies but are un-chair-like), we learn a \emph{mapping network} $F$ to transform the generator's input latent code $\mathbf{z}$ to a latent code $F(\mathbf{z})$.
$F$ induces a body-conditional warping of the latent space, pushing latent codes towards regions of the space which better accommodate the input body.
This approach is based on prior work that fine-tunes 3D shape generative models to be physically connected and stable~\cite{Mezghanni_2021_CVPR}.

Figure~\ref{fig:gen_condition} illustrates this approach.
Given an input body shape or pose, we project it to a conditioning vector $\mathbf{c}$, which is then fed to a FiLM network.
The FiLM net produces element-wise scales and shifts for each layer of the mapping network $F$, which takes a latent code $\mathbf{z}$ as input and produces a transformed latent code $F(\mathbf{z})$ as output.
We train two variants of $F$: one that takes body shapes as input (18-dimensional SMPL shape vector), and one that takes target poses (69-dimensional SMPL pose vectors).
$F(\mathbf{z})$  is then fed to the fixed generator network to produce an output shape, which is then point-sampled and fed to the appropriate loss proxy network to produce a training loss.
We regularize the network with an additional loss that penalizes $F(\mathbf{z})$ from drifting too far from $\mathbf{z}$.

We use this procedure to train shape- and pose-conditioned variants for three generative shape models: ShapeAssembly, a generative model that writes programs which declare and connect cuboids~\cite{13}; SP-GAN, a point cloud generator~\cite{6}; and IM-Net, an implicit field generator~\cite{7}.
More training details can be found in supplemental.
\section{Results \& Evaluation}
\label{sec:results}

In this section, we evaluate the performance of each component of our system: chair functionality metrics, loss proxy networks, and body-conditional generative models.

\parahead{Comfort metric perceptual validation}
To evaluate how well the comfort metric proposed in Section~\ref{sec:comfort_metric} agrees with human judgments of comfort, we conduct a two-alternative forced choice perceptual study.
In each comparison, participants were shown two images of a human body sitting in a chair and asked which sitting pose looks more comfortable.
We recruited 20 participants (members of our research lab), each of whom performed the same set of 50 such comparisons.
The relative values of our comfort metric agree with the participants for $70\%$ of these comparisons (95\% confidence interval: $(55\%, 86\%)$), indicating that our metric generally captures human perception of sitting comfort.

\parahead{Accuracy of loss proxy networks}
We evaluate how well our two loss proxy networks learn to approximate the true loss functions by checking whether they order chairs by loss in the same way.
Specifically, we create a large set of pairs of chairs (not seen during training) and record which of the pair has a lower true loss value. 
We then check how often the chair they predict the lower loss for is the same one that has the lower true loss.
Table~\ref{tab:proxy_acc} reports these accuracies.
The comfort loss proxy is harder to learn than the pose proxy, as comfort is the more subtle of the two objectives.
As we will show, this level of accuracy is sufficient for effectively fine-tuning our generative models.

\begin{table}[t!]
    \centering
    \footnotesize
    \renewcommand{\arraystretch}{0.85}
    \begin{tabular}{lcc}
        \toprule
        \textbf{Data type} & \textbf{Comfort Proxy} & \textbf{Pose Proxy}
        \\
        \midrule
        Cuboids (ShapeAssembly) & 64\% & 89\% 
        \\
        Implicit (IM-Net) & 62\% & 80\%
        \\
        Point cloud (SP-GAN) & 73\% & 88\%
        \\
        \bottomrule
    \end{tabular}
    \vspace{-1em}
    \caption{
    Accuracy of our loss proxy networks when used to predict which of two chairs should have the lower loss.}
    \label{tab:proxy_acc}
\end{table}

\parahead{Generative model evaluation}
We evaluate our body-conditional generative models using the following metrics:
\begin{packed_itemize}
    \item \emph{Comfort loss (Comfort)}: the mean comfort loss value for a chair generated given an input body shape (in kilopascals; lower is better).
    \item \emph{Poss loss (Pose)}: the mean pose matching loss value for a chair generated given an input body shape (in centimeters; lower is better).
    \item \emph{Frechet Distance (FD)}: the Frechet Distance~\cite{FrechetInceptionDistance} (evaluated in the feature space of a pre-trained PointNet classifier~\cite{qi2017pointnet}) between a set of generated chairs and a set of chairs from a held-out test set (lower is better).
\nolistbottomspace
\end{packed_itemize}
We compare to the original generative model pre-trained on the PartNet dataset as well as an ``oracle'' method in which multiple (10) samples are drawn from the original generative model and then evaluated using our comfort loss or pose matching loss; whichever achieves the best loss is returned as the output of this method.
The oracle requires running the expensive sitting pose optimization for each chair, so it is not practical for most applications.
It does serve as a informative upper bound on performance, however.

Table~\ref{tab:gen_quant} shows the results of this experiment.
The body-conditioned variants achieve better pose and comfort losses than the original model (though not quite as good as the expensive-to-evaluate oracle).
Respecting the functional losses comes at the cost of a small distribution shift (as measured by FD), which the oracle also incurs.
Figure~\ref{fig:comfort_qual} qualitatively compares some outputs of our shape-conditioned generative model with that of the original generative model when given the same latent code; Figure~\ref{fig:target_qual} shows analogous results for our pose-conditioned generative model.
The fine-tuned generative models adjust chair geometry to better accommodate the input body: tilting chair backs forward or back, lowering or raising seat slope, etc.
Our regularization that penalizes the warped latent code from drifting too far from the original latent code does prevent some large-scale geometric changes from happening (e.g. adding footrests to the chair in Figure~\ref{fig:target_qual} bottom, second from left).

\begin{table}[t!]
    \centering
    \footnotesize
    \setlength{\tabcolsep}{3pt}
    \renewcommand{\arraystretch}{0.85}
    \begin{tabular}{llccc}
        \toprule
        \textbf{Model type} & \textbf{Model variant} &
        \textbf{Comfort$\downarrow$} & \textbf{Pose$\downarrow$} &
        \textbf{FD$\downarrow$}
        \\
        \midrule
        \multirow{5}{*}{ShapeAssembly} &
        Original & 11.2 & 44.5 & 33.0 \\   
        & Shape-conditioned & 7.71 & -- & 41.8\\
        & Pose-conditioned & -- & 30.3 & 42.8\\
        & \emph{Oracle} & \emph{6.52} & \emph{21.2} & \emph{39.3}
        \\
        \midrule
        \multirow{5}{*}{IM-Net} &
        Original & 14.1 & 40.35 & 82.6 \\
        & Shape-conditioned & 10.7 & -- & 86.0\\
        & Pose-conditioned & -- & 31.6 & 92.0\\
        & \emph{Oracle} & \emph{9.53} & \emph{17.25} & \emph{86.8}
        \\
        \midrule
        \multirow{5}{*}{SP-GAN} &
        Original & 10.23 & 109.85 & 25.4 \\  
        & Shape-conditioned & 8.77 & -- & 67.1\\
        & Pose-conditioned & -- & 61.87 & 60.5\\
        & \emph{Oracle} & \emph{4.74} & \emph{31.04} & \emph{25.1} 
        \\
        \bottomrule
    \end{tabular}
    \vspace{-1em}
    \caption{Evaluating how well different generative models respect functional losses (Comfort, Pose) while staying within the distribution of chairs (FD).
    }
    \label{tab:gen_quant}
\end{table}

\begin{figure*}[t!]
    \centering
    \small
    \setlength{\tabcolsep}{1pt}
    \renewcommand{\arraystretch}{0.5}
    \begin{tabular}{c cc cc cc}
        & \multicolumn{2}{c}{ShapeAssembly} & \multicolumn{2}{c}{IM-Net} & \multicolumn{2}{c}{SP-GAN}
        \\
        \raisebox{0.8em}{\rotatebox{90}{Unconditioned}} &
        \includegraphics[width=0.16\linewidth]{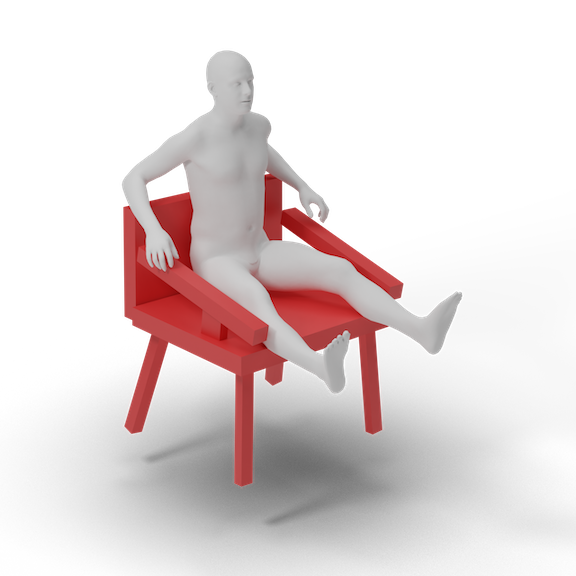} &
        \includegraphics[width=0.16\linewidth]{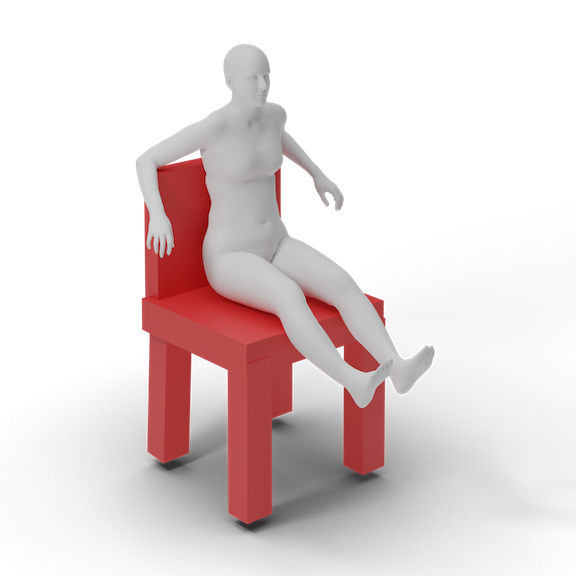} &
        \includegraphics[width=0.16\linewidth]{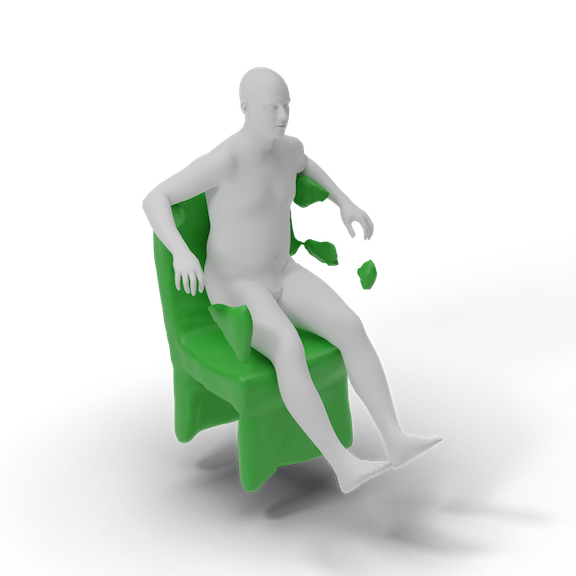} &
        \includegraphics[width=0.16\linewidth]{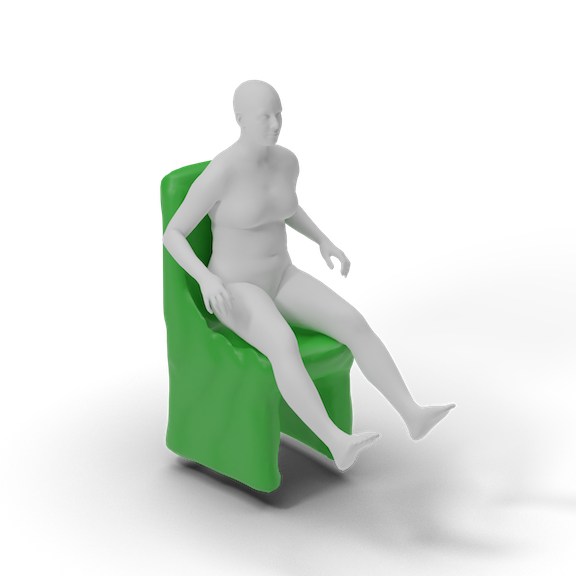} &
        \includegraphics[width=0.16\linewidth]{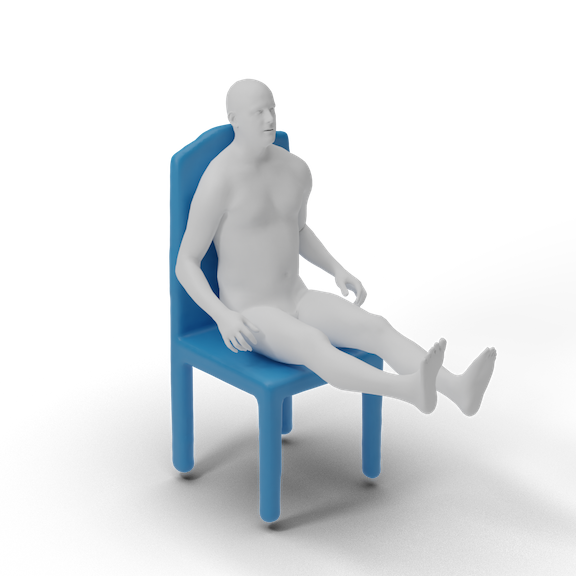} &
        \includegraphics[width=0.16\linewidth]{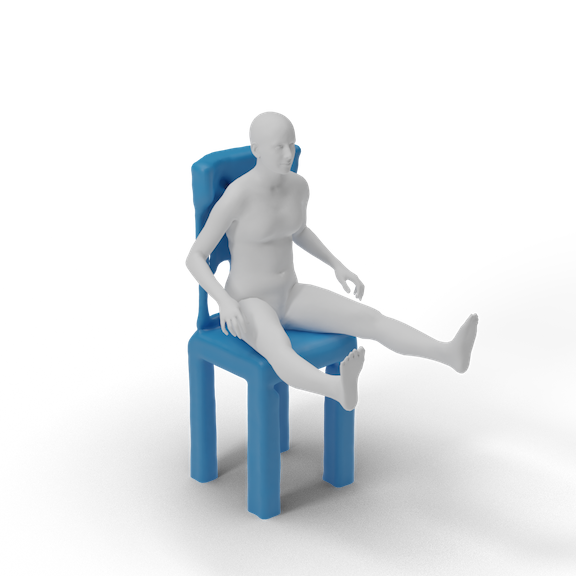} \\
        & $\mathcal{L}_\text{comfort} = 11.0$ &
        $\mathcal{L}_\text{comfort} = 11.8$ &
        $\mathcal{L}_\text{comfort} = 13.5$ &
        $\mathcal{L}_\text{comfort} = 14.3$ &
        $\mathcal{L}_\text{comfort} = 11.2$&
        $\mathcal{L}_\text{comfort} = 10.9$
        \\
        \raisebox{1.3em}{\rotatebox{90}{Conditioned}} &
        \includegraphics[width=0.16\linewidth]{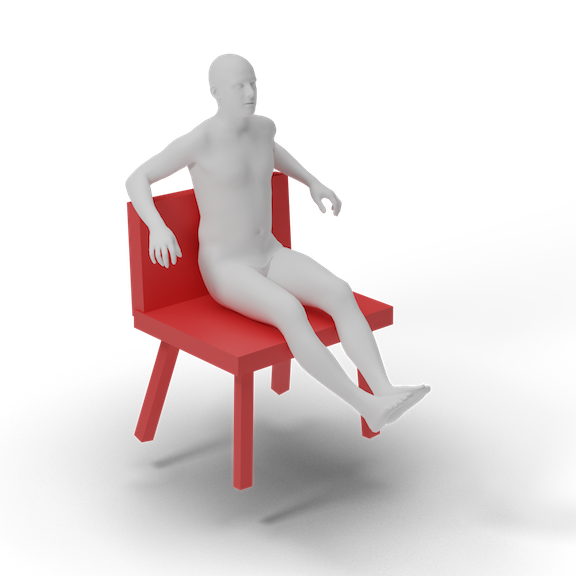} &
        \includegraphics[width=0.16\linewidth]{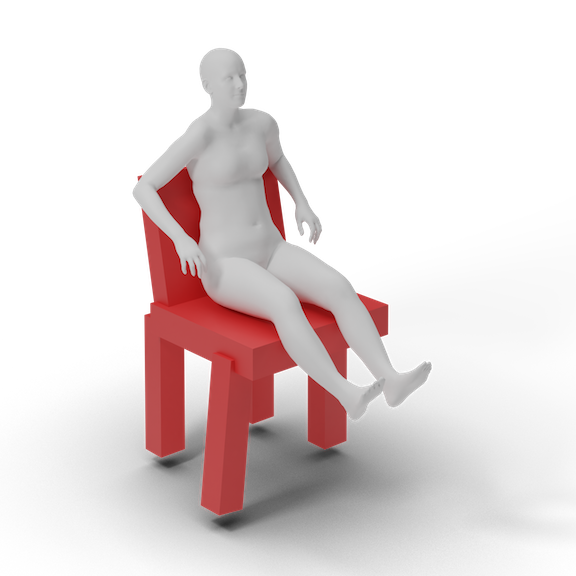} &
        \includegraphics[width=0.16\linewidth]{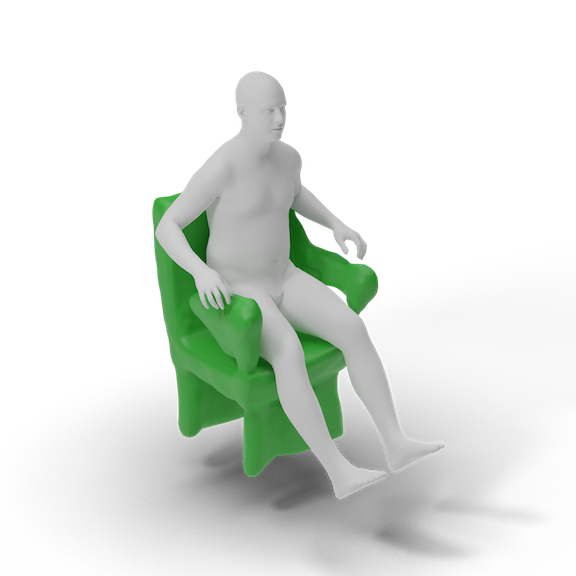} &
        \includegraphics[width=0.16\linewidth]{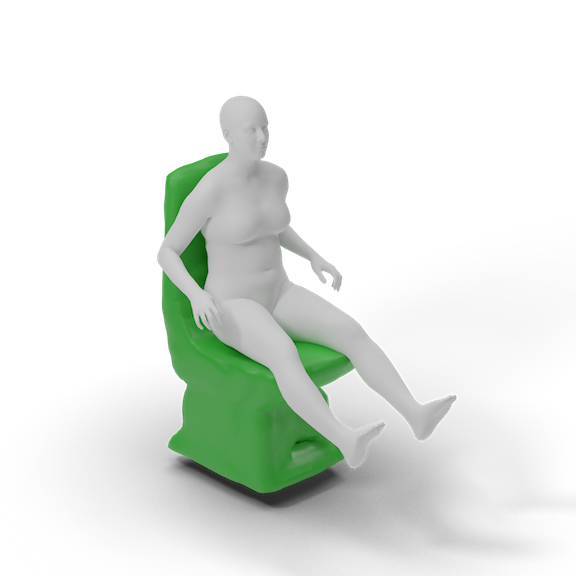} &
        \includegraphics[width=0.16\linewidth]{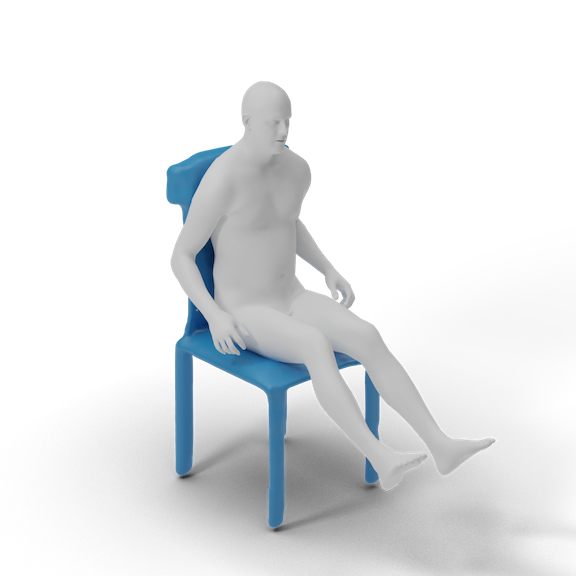} &
        \includegraphics[width=0.16\linewidth]{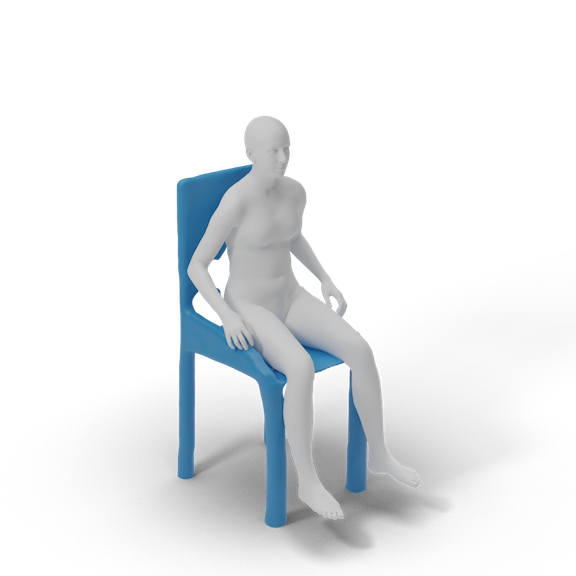}\\
        & $\mathcal{L}_\text{comfort} = 7.93$ &
        $\mathcal{L}_\text{comfort} = 7.07$ &
        $\mathcal{L}_\text{comfort} = 11.3$ &
        $\mathcal{L}_\text{comfort} = 10.5$ &
        $\mathcal{L}_\text{comfort} = 9.90$ &
        $\mathcal{L}_\text{comfort} = 8.06$
    \end{tabular}
    \caption{Comparing outputs of our shape-conditioned generative models with their corresponding unconditioned generative models, given the same latent code.}
    \label{fig:comfort_qual}
\end{figure*}

\begin{figure*}[t!]
    \centering
    \small
    \setlength{\tabcolsep}{1pt}
    \renewcommand{\arraystretch}{0.5}
    \begin{tabular}{c cc cc cc}
        & \multicolumn{2}{c}{ShapeAssembly} & \multicolumn{2}{c}{IM-Net} & \multicolumn{2}{c}{SP-GAN}
        \\
        \raisebox{0.8em}{\rotatebox{90}{Unconditioned}} &
        \includegraphics[width=0.16\linewidth]{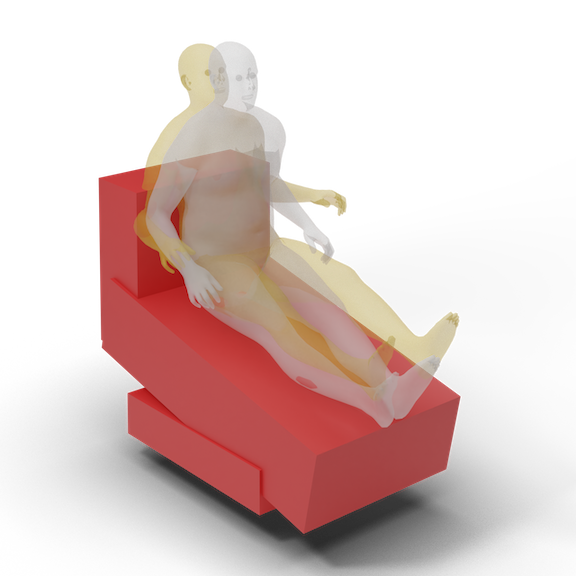} &
        \includegraphics[width=0.16\linewidth]{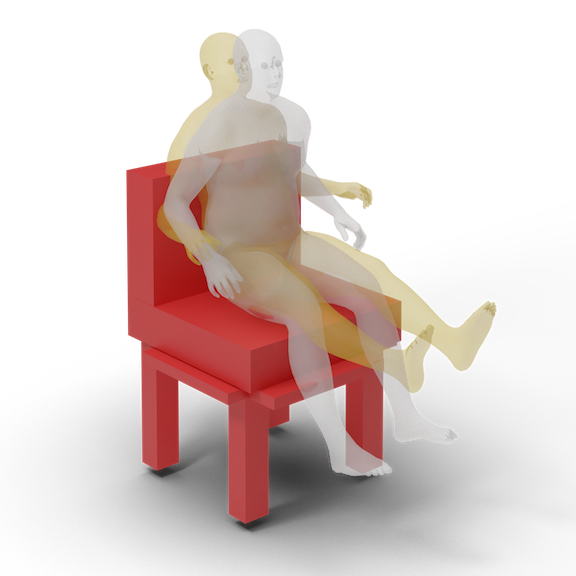} &
        \includegraphics[width=0.16\linewidth]{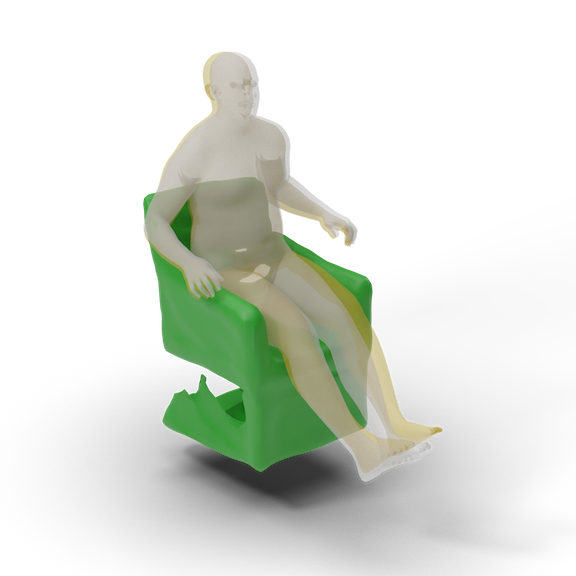} &
        \includegraphics[width=0.16\linewidth]{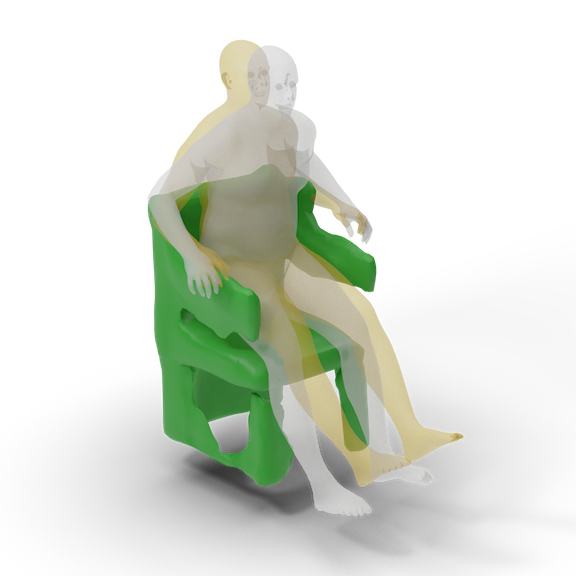} &
        \includegraphics[width=0.16\linewidth]{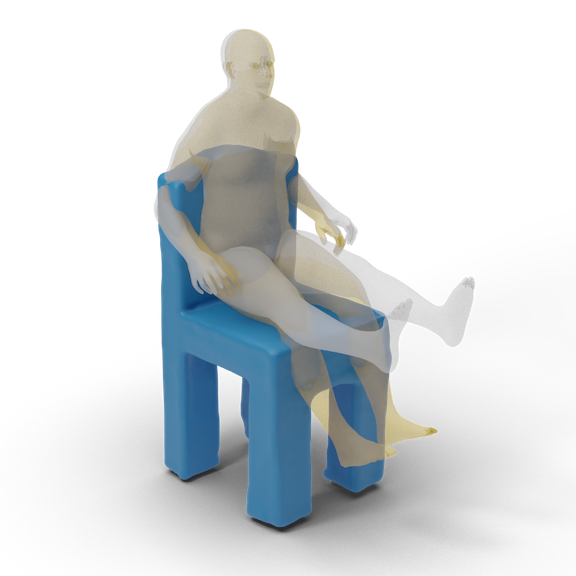} &
        \includegraphics[width=0.16\linewidth]{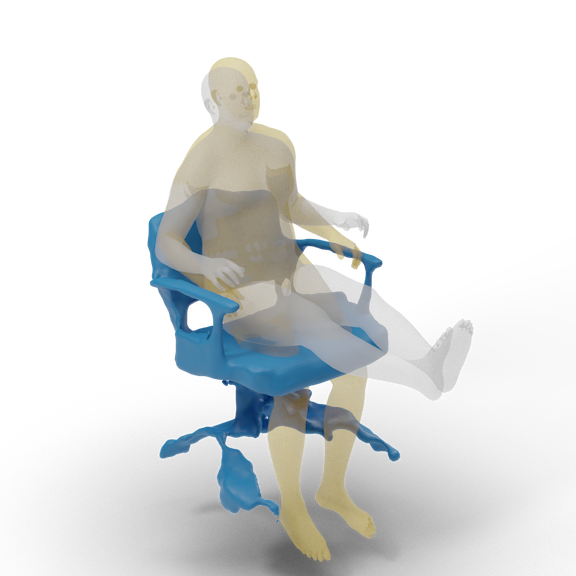}\\
        & $\mathcal{L}_\text{pose} = 17.2$ &
        $\mathcal{L}_\text{pose} = 49.2$ &
        $\mathcal{L}_\text{pose} = 16.8$ &
        $\mathcal{L}_\text{pose} = 45.6$ &
        $\mathcal{L}_\text{pose} = 31.5$ &
        $\mathcal{L}_\text{pose} = 84.0$
        \\
        \raisebox{1.3em}{\rotatebox{90}{Conditioned}} &
        \includegraphics[width=0.16\linewidth]{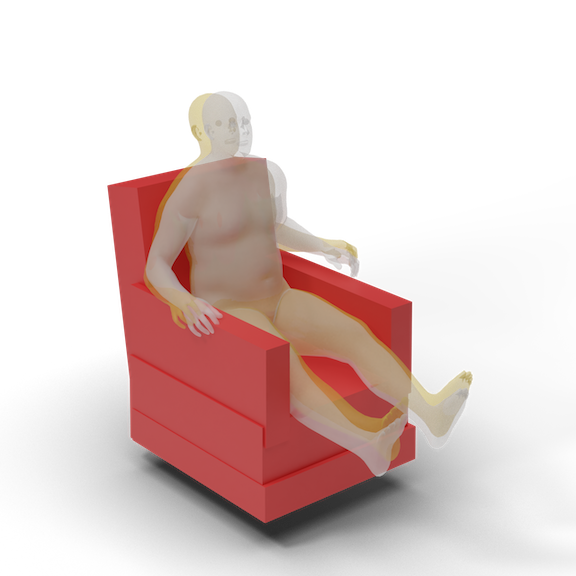} &
        \includegraphics[width=0.16\linewidth]{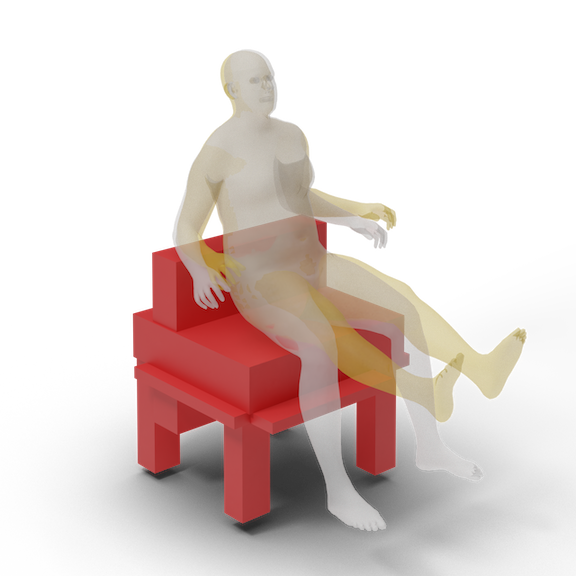} &
        \includegraphics[width=0.16\linewidth]{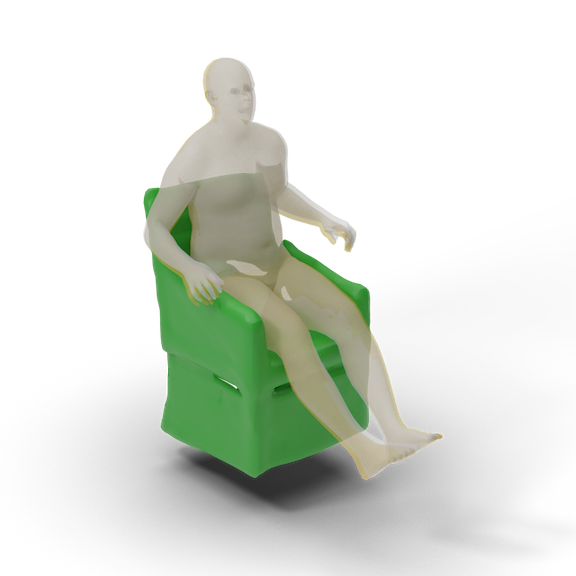} &
        \includegraphics[width=0.16\linewidth]{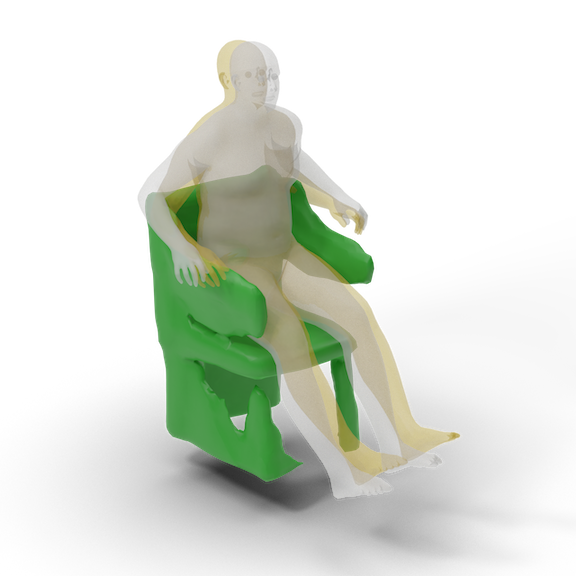} &
        \includegraphics[width=0.16\linewidth]{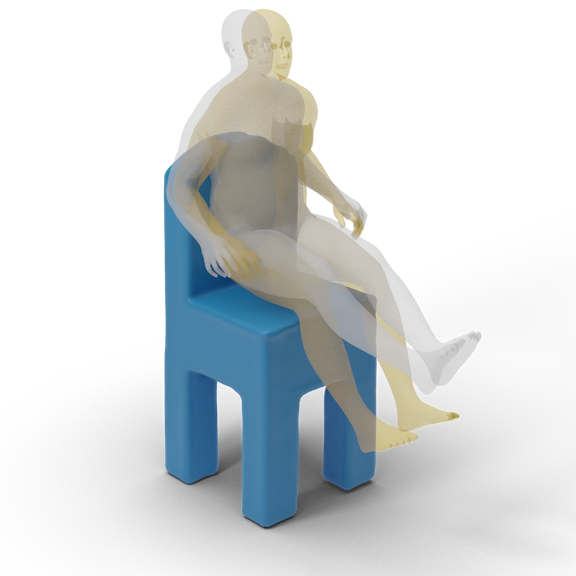} &
        \includegraphics[width=0.16\linewidth]{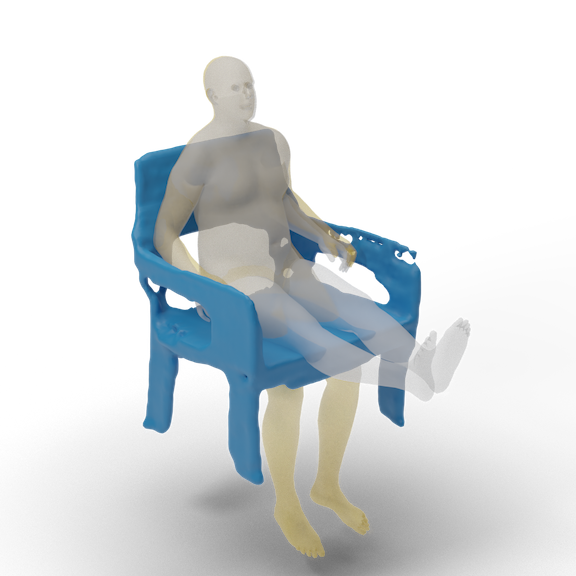}\\
        & $\mathcal{L}_\text{pose} = 8.40$ &
        $\mathcal{L}_\text{pose} = 15.5$ &
        $\mathcal{L}_\text{pose} = 8.25$ &
        $\mathcal{L}_\text{pose} = 24.1$ &
        $\mathcal{L}_\text{pose} = 14.2$ &
        $\mathcal{L}_\text{pose} = 69.9$
    \end{tabular}
    \caption{Comparing outputs of our pose-conditioned generative models with their corresponding unconditioned generative models, given the same latent code. The target pose is shown in gold; the pose the body actually assumes when sitting in the chair is shown in gray.}
    \label{fig:target_qual}
\end{figure*}
\section{Conclusion}
\label{sec:conclusion}
We presented a new technique for adapting generative models of 3D chairs into \emph{body-aware} generative models which accommodate input body shapes or sitting postures.
We described an optimization for finding a body's sitting pose in a chair, defined a metric for assessing the comfort of such sitting poses, and trained neural networks to approximate this metric and a related pose-matching metric.
Finally, we introduced a general scheme for adding body conditioning to any latent variable shape generative model and trained these models to minimize network proxy losses.
The generality of this scheme allows application to three base generative models: ShapeAssembly, IM-Net, and SP-GAN.

\parahead{Limitations \& Future Work}
While our method is currently limited to chairs, we look forward to learning to synthesize other common objects that people interact (e.g.~other furniture, tools).
Our sitting pose optimization procedure and comfort loss metric use limited physics and could benefit from soft body simulation to produce more accurate body pressure estimates.
We could also remove the assumption that the sitting pose is passive, incorporating active human agents that seek to accomplish goals while sitting (e.g.~eating, working at a desk, or conversing with a friend).
We could replace the current sitting pose optimization procedure with one that factors in the active effort required for a person to sit.
Pursuing this latter direction would facilitate design of chairs for people with limited strength, mobility, or muscle control.
There is no reason to limit the model to working on the range of body shapes expressible in the SMPL model: it would be valuable to support the bodies of children, or people missing one or more limbs.

{\small
\bibliographystyle{ieee_fullname}
\bibliography{bib.bib}
}

\appendix
\section{Supplemental Materials}
\subsection{Loss Proxy Training}
We found that the ground-truth comfort loss and the pose matching loss distributions were too widely dispersed for a neural network to learn to regress. We found that by taking the log of the comfort loss and the log of the pose matching loss increased by 1, the loss distributions better resembled non-unit normal distributions. We then whitened these values to have a mean of zero and a standard deviation of one.
\subsection{Generative Model Training}
We found that when training our conditioning networks to produce latent vectors for our generative models, we needed to regularize the values produced by the network to stop the models from exceeding too far beyond the unit normal distribution and creating shapes that stop functioning as chairs.
\subsubsection{Shape Assembly}
When training ShapeAssembly we found that leaving the model to train without a regularizer manifests in the generation of erroneous shape programs, as only a subset of programs output by the decoder are executable. We implemented a $\mathbf{z}$-distance loss with learning weight 0.001 to restrict the model from producing latent vectors far from the unconditioned vector. We train the ShapeAssembly conditioning network for 50 epochs, 1000 iterations for each epoch. We use a learning rate of 0.0001 with no learning rate scheduling. 
\subsubsection{IM-NET}
For IM-NET, training without regularization yields chairs with artefacts or signed distance function outputs that do not contain the level set at which we produce our output chair meshes through the marching cubes algorithm. Similarly, we implemented a $\mathbf{z}$-distance loss with learning weight 0.001 to restrict the model from producing errors. For SP-GAN, we likewise found that training without any regularization term results in deformed shapes that do not resemble chairs. Thus, we implemented a $\mathbf{z}$-distance loss with coefficient $0.1$. We train the IM-NET conditioning network for 50 epochs, 1000 iterations for each epoch. We use a learning rate of 0.0001 with no learning rate scheduling. 
\subsubsection{SP-GAN}
The SP-GAN decoding process is slightly different, as the model decodes a chair from $N_p$ $\mathbf{z}$ vectors where $N_p$ is the number of points on the output pointcloud. This gives the output space significantly higher degrees of freedom, so we can only train the model for very few iterations before the output shapes start to deform. Thus, we only need to train SP-GAN with learning rate 0.0001 with no learning rate scheduling for 100 epochs for 10 iterations per epoch in order to see results. Due to the limited amount of training, the various training setups for SP-GAN's pose models do not result in large $\mathbf{z}$ distance deviations. However, we still found that adding in the $\mathbf{z}$ distance loss with coefficient $0.1$ allowed our model to produce lower loss values from our pose and comfort loss proxies. 

\end{document}